\documentclass[
aps,%
12pt,%
final,%
notitlepage,%
oneside,%
onecolumn,%
nobibnotes,%
nofootinbib,%
superscriptaddress,%
noshowpacs]%
{revtex4}
\usepackage{graphicx}
\usepackage{color}

\begin{document}

\noindent {\it Astronomy Reports, 2026, Vol. , No. }
\bigskip\bigskip  \hrule\smallskip\hrule
\vspace{35mm}

\title{BLACK HOLES IN ALTERNATIVE THEORIES OF GRAVITY\footnote{Paper presented at the Sixth Zeldovich meeting, an international conference in honor of Ya. B. Zeldovich held in Pescara, Italy on July 13--17, 2026. Published by the recommendation of the special editors: R. Ruffini and G. V. Vereshchagin.}}

\author{\bf \copyright $\:$ 2026.
\quad
\firstname{J.~L.}~\surname{Bl\'azquez-Salcedo}}%
\email{jlblaz01@ucm.es}
\affiliation{Departamento de F\'isica Te\'orica and IPARCOS, Facultad de Ciencias F\'isicas, Universidad Complutense de Madrid, Spain}%

\author{\bf \firstname{F.~S.}~\surname{Khoo}}%
\email{fkhoo@ucm.es}
\affiliation{Departamento de F\'isica Te\'orica and IPARCOS, Facultad de Ciencias F\'isicas, Universidad Complutense de Madrid, Spain}%

\author{\bf \firstname{B.}~\surname{Kleihaus}}%
\email{b.kleihaus@uni-oldenburg.de}
\affiliation{Institute of Physics, University of Oldenburg, Oldenburg, Germany}%

\author{\bf  \firstname{J.}~\surname{Kunz}}%
\email{jutta.kunz@uni-oldenburg.de}
\affiliation{Institute of Physics, University of Oldenburg, Oldenburg, Germany}%

\begin{abstract}

\centerline{\footnotesize Received: ;$\;$
Revised: ;$\;$ Accepted: .}\bigskip\bigskip\bigskip

Black holes, with their strong gravitational fields, provide an important testing ground for theories of gravity beyond General Relativity. 
Among the many proposed alternatives, considerable recent work has focused on scalar-tensor theories in which a scalar field couples to higher-curvature terms. 
The black hole solutions that arise in such theories can differ significantly from the Schwarzschild and Kerr solutions of General Relativity. 
Characteristic properties of these black holes include instabilities, shadows, and gravitational wave spectra, which can be used to constrain the couplings of the underlying theories.

\end{abstract}

\maketitle

\section{Introduction}

The vacuum black holes of General Relativity (GR) are the static spherically symmetric Schwarzschild black holes and the stationary axially symmetric Kerr black holes.
The rotating Kerr black holes are specified completely by only two parameters, their mass $M$ and their angular momentum $J$ \cite{Chrusciel:2012jk}, and the Kerr bound $J/M^2\leq1$ limits their domain of existence.
Kerr black holes are of overwhelming importance, since the exterior of astrophysical black holes is expected to be well described by the Kerr solution.
This expectation is often referred to as the \textit{Kerr paradigm} \cite{Bambi:2015awa}.
This paradigm is now being tested by the observations of black hole shadows \cite{EventHorizonTelescope:2019dse,EventHorizonTelescope:2022wkp} and, in particular, by the detection of gravitational waves from black hole mergers \cite{LIGOScientific:2016aoc}.

While all current observations are compatible with GR predictions \cite{Will:2018bme}, there is a variety of serious reasons to explore alternative theories of gravity like the unknown nature of dark matter and dark energy, and our limited understanding of gravity at the quantum level.
Thus, numerous theories of gravity have been proposed and investigated in the weak- and strong-gravity regimes and in cosmology \cite{Berti:2015itd,CANTATA:2021ktz}.
An interesting set of these theories contains additional scalar degrees of freedom, which may couple non-minimally to curvature invariants and allow black holes to carry scalar hair.
A particularly well-studied class is formed by the Einstein-scalar-Gauss-Bonnet (EsGB) theories (for a recent review see \cite{Doneva:2022ewd}).

Here, we review several recent developments for these scalarized black holes.
Our main emphasis is on Einstein-dilaton-Gauss-Bonnet (EdGB) black holes and the recently obtained quasinormal-mode (QNM) spectra of rapidly rotating solutions.
We then turn to curvature-induced spontaneous scalarization, where rotation can suppress or induce scalar hair, and conclude with a brief discussion of Einstein-scalar-Gauss-Bonnet-Ricci (EsGBR) black holes and their stability.

\section{Einstein-scalar-Gauss-Bonnet theories}

The action of EsGB gravity may be written as
\begin{equation}
S=\frac{1}{16\pi}\int d^4x\sqrt{-g}
\left[R-\frac{1}{2}(\partial_\mu\varphi)^2+f(\varphi)R^2_{\rm GB}\right],
\label{action}
\end{equation}
where
\begin{equation}
R^2_{\rm GB}=R_{\mu\nu\rho\sigma}R^{\mu\nu\rho\sigma}
-4R_{\mu\nu}R^{\mu\nu}+R^2 .
\end{equation}
Theories of this type yield second-order field equations and form a sector of Horndeski theory \cite{Horndeski:1974wa}.
The scalar equation,
\begin{equation}
\nabla_\mu\nabla^\mu\varphi+\frac{df}{d\varphi}R^2_{\rm GB}=0 ,
\label{scalar}
\end{equation}
shows immediately that the choice of the coupling function $f(\varphi)$ is of utmost importance, since two qualitatively different situations arise. 
For coupling functions satisfying $f'(0)\ne 0$, like the dilatonic coupling motivated by the low-energy limit of string theory \cite{Gross:1986mw,Metsaev:1987zx}, the GR black holes are no longer solutions of the field equations, and only hairy black holes are present. 
For coupling functions satisfying $f'(0)=0$, on the other hand, the GR black holes do remain solutions, but additional scalarized black holes can emerge.

\subsection{Einstein-dilaton-Gauss-Bonnet black holes}

For the dilatonic case, we choose
\begin{equation}
f(\varphi)=\frac{\alpha}{4}e^{-\gamma\varphi},
\end{equation}
with the Gauss-Bonnet coupling $\alpha$ and the string theory motivated value $\gamma=1$.
Static EdGB black holes were first obtained in Ref.~\cite{Kanti:1995vq}.
For fixed coupling $\alpha$, they possess a lower bound on the mass.
Here, the fundamental branch reaches a critical configuration.
This branch is stable against radial perturbations, and its QNM spectrum shows that the axial-polar isospectrality of Schwarzschild black holes is broken by the presence of the scalar, and that additional scalar-led modes arise \cite{Blazquez-Salcedo:2016enn,Blazquez-Salcedo:2017txk}.

\begin{figure}
\begin{center}
\includegraphics[width=0.72\textwidth,angle=0]{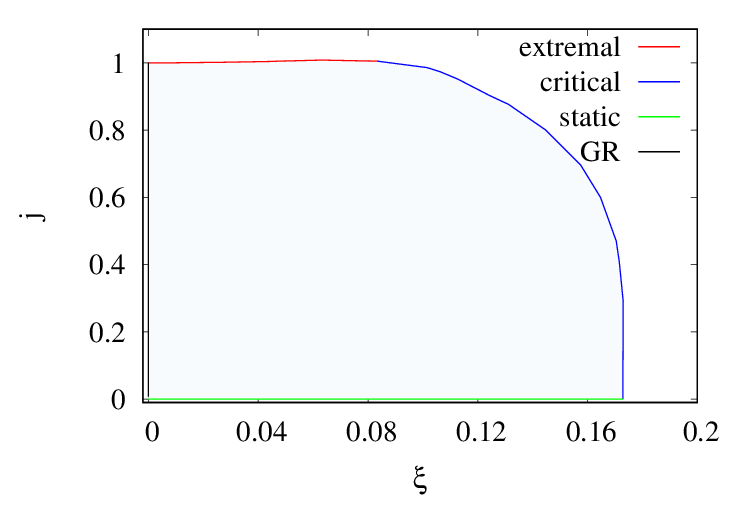}
\end{center}
\caption{Domain of existence of rotating EdGB black holes: scaled angular momentum
$j=J/M^2$ versus scaled coupling constant $\xi=\alpha/M^2$.
The domain is bounded by static, critical, extremal, and Kerr solutions.}
\label{Fig1}
\end{figure}

The static solutions can be extended to include rotation \cite{Kleihaus:2011tg}.
The domain of existence of the rotating black holes is bounded by static, critical, extremal, and Kerr black holes, as illustrated in Fig.~\ref{Fig1}, where the dimensionless angular momentum $j=J/M^2$ is shown versus the dimensionless coupling $\xi=\alpha/M^2$.
For $0\leq j\lesssim0.286$ the maximal coupling is approximately $\xi_{\rm max}=0.173$, becoming smaller for more rapid rotation.
Interestingly, the solutions can slightly exceed the Kerr bound.
Moreover, they possess multipole moments, orbital properties, and shadows that differ (at least slightly) from those of Kerr black holes with the same global charges \cite{Kleihaus:2015aje,Cunha:2016wzk}.

Gravitational waves provide a particularly promising way of probing these deviations.
An inspiral-merger-ringdown waveform has recently been constructed in EdGB gravity within the effective-one-body formalism \cite{Julie:2024fwy}.
The inspiral part of their analysis already established a new bound for the coupling.
The late-time signal is governed by the QNM spectrum of the final black hole, and it is therefore important to determine this spectrum also for rapidly rotating backgrounds.

\subsection{Quasinormal modes of rapidly rotating EdGB black holes}

We linearly perturb the rotating black hole background ($g^{(0)}_{\mu\nu}$, $\varphi_0$) 
\begin{eqnarray}
g_{\mu\nu}&=&g^{(0)}_{\mu\nu}+\epsilon h_{\mu\nu},\nonumber\\
\varphi&=&\varphi_0+\epsilon\,\delta\varphi ,
\end{eqnarray}
where $\epsilon$ is the perturbation parameter.
We then factor the dependence on time and on the azimuthal angle as $\exp[i(M_z\phi-\omega t)]$ with azimuthal number $M_z$.
The complex eigenfrequency
\begin{equation}
\omega=\omega_R+i\omega_I
\end{equation}
determines the oscillation frequency and the damping time $\tau=1/|\omega_I|$.
For rotating EdGB black holes, the metric and scalar perturbations are coupled, and the perturbation equations form a system of partial differential equations in the radial and polar coordinates.

Whereas earlier calculations employed expansions in the rotation and the coupling strength \cite{Pierini:2021jxd,Pierini:2022eim}, we recently obtained the QNMs on fully non-perturbative rapidly rotating EdGB backgrounds \cite{Blazquez-Salcedo:2024oek,Blazquez-Salcedo:2025spectrum}.
We employed a spectral decomposition of the perturbation functions in Chebyshev polynomials in the radial coordinate and associated Legendre functions in the polar coordinate, leading to a quadratic generalized eigenvalue problem, that was solved numerically.
We classified the modes as polar-led, axial-led, and scalar-led, according to their Kerr limits for vanishing Gauss-Bonnet coupling.
Then the polar-led and axial-led modes approach the isospectral Kerr modes, whereas the scalar-led modes reduce to the scalar test-field modes.

\begin{figure}
\begin{center}
\includegraphics[width=1.0\textwidth,angle=-90]{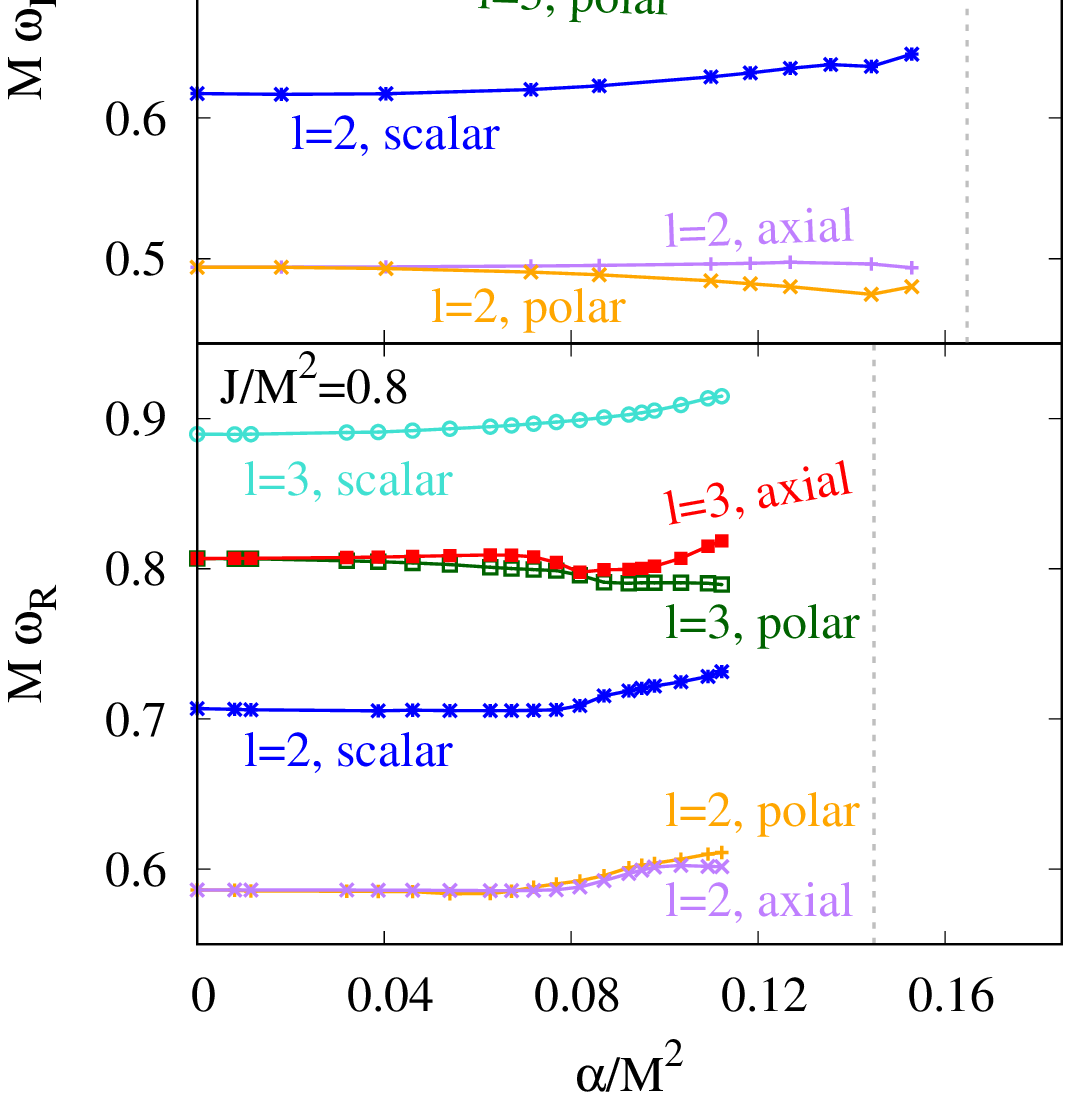}
\includegraphics[width=1.0\textwidth,angle=-90]{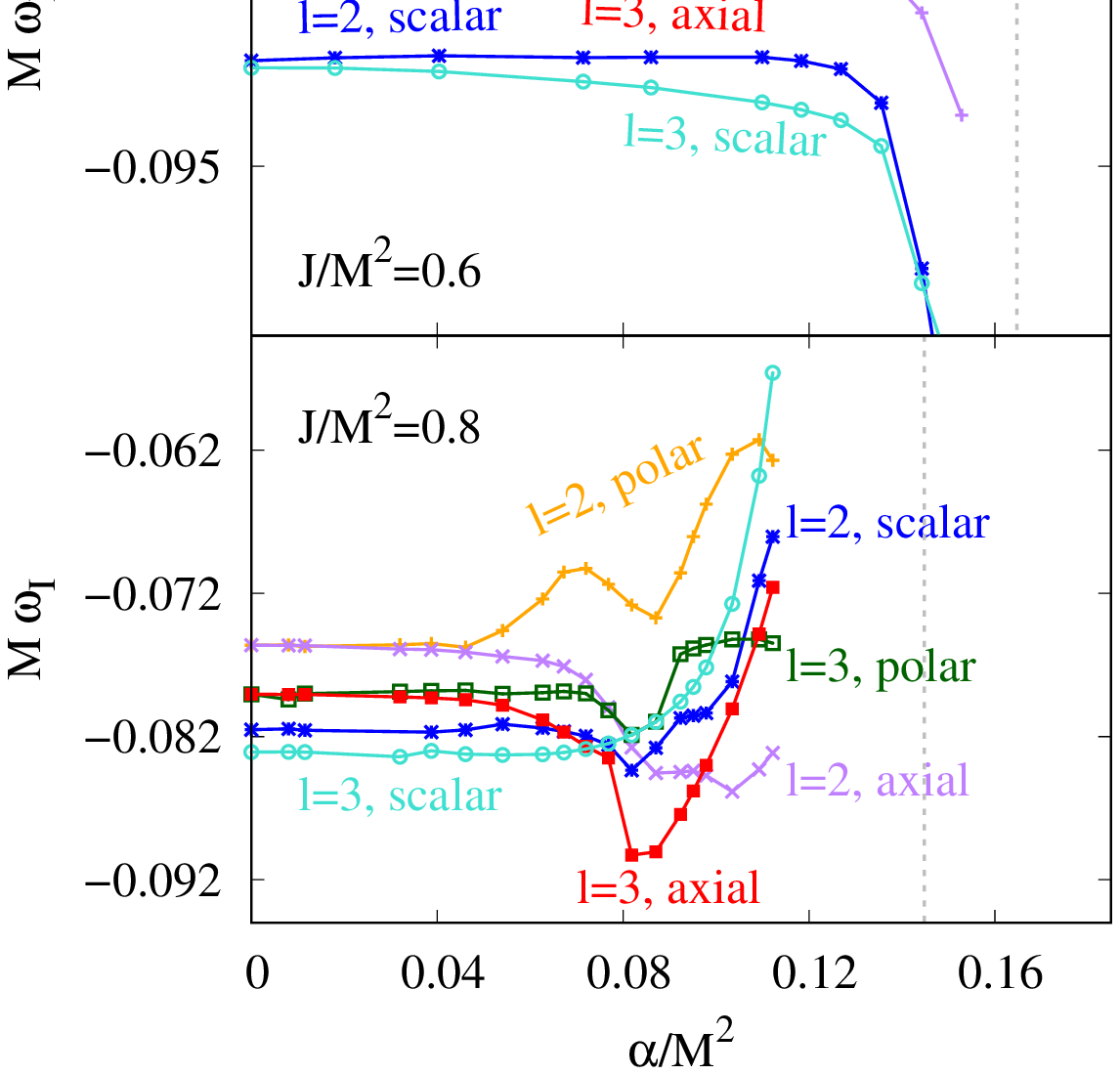}
\end{center}
\caption{Fundamental EdGB $(l=2)$-led and $(l=3)$-led QNMs for $M_z=2$.
The scaled real part $M\omega_R$ (left) and scaled imaginary part $M\omega_I$ (right)
are shown versus the scaled coupling strength $\xi$ for $j=0.2$, 0.4, 0.6, and 0.8.
Each multipole features distinct polar-led, axial-led, and scalar-led modes.}
\label{Fig2}
\end{figure}

Figure~\ref{Fig2} shows the fundamental $(l=2)$-led and $(l=3)$-led QNMs for $M_z=2$ and four values of the angular momentum, $j=0.2$, 0.4, 0.6, and 0.8.
The Kerr isospectrality is broken when the Gauss-Bonnet coupling is switched on, and the splitting typically increases toward the boundary of the domain of existence.
The real parts change comparatively smoothly, whereas the damping times can depend strongly on both the coupling and the angular momentum.
Consequently, the longest-lived mode can change its character as the parameters are varied.
For small $j$, an axial-led quadrupole mode dominates at small coupling, while close to the maximum coupling, a scalar-led mode can become the longest-lived mode.
At larger rotation, the damping rates of the different mode sectors exhibit increasingly distinct dependencies on the coupling.
We found no unstable fundamental modes in the investigated domain.

\begin{figure}
\begin{center}
\mbox{ 
\includegraphics[width=0.36\textwidth,angle=-90]{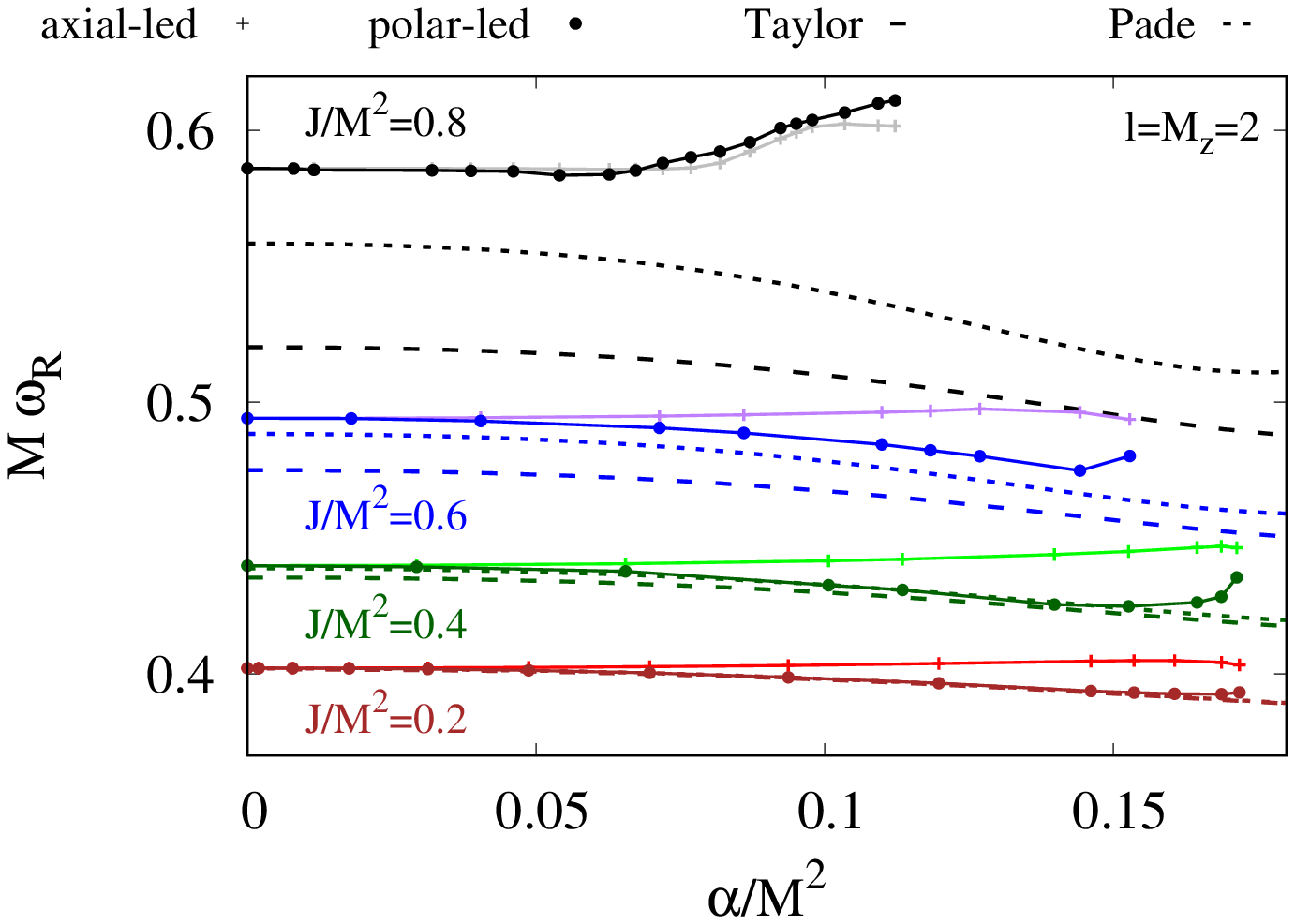}
\includegraphics[width=0.36\textwidth,angle=-90]{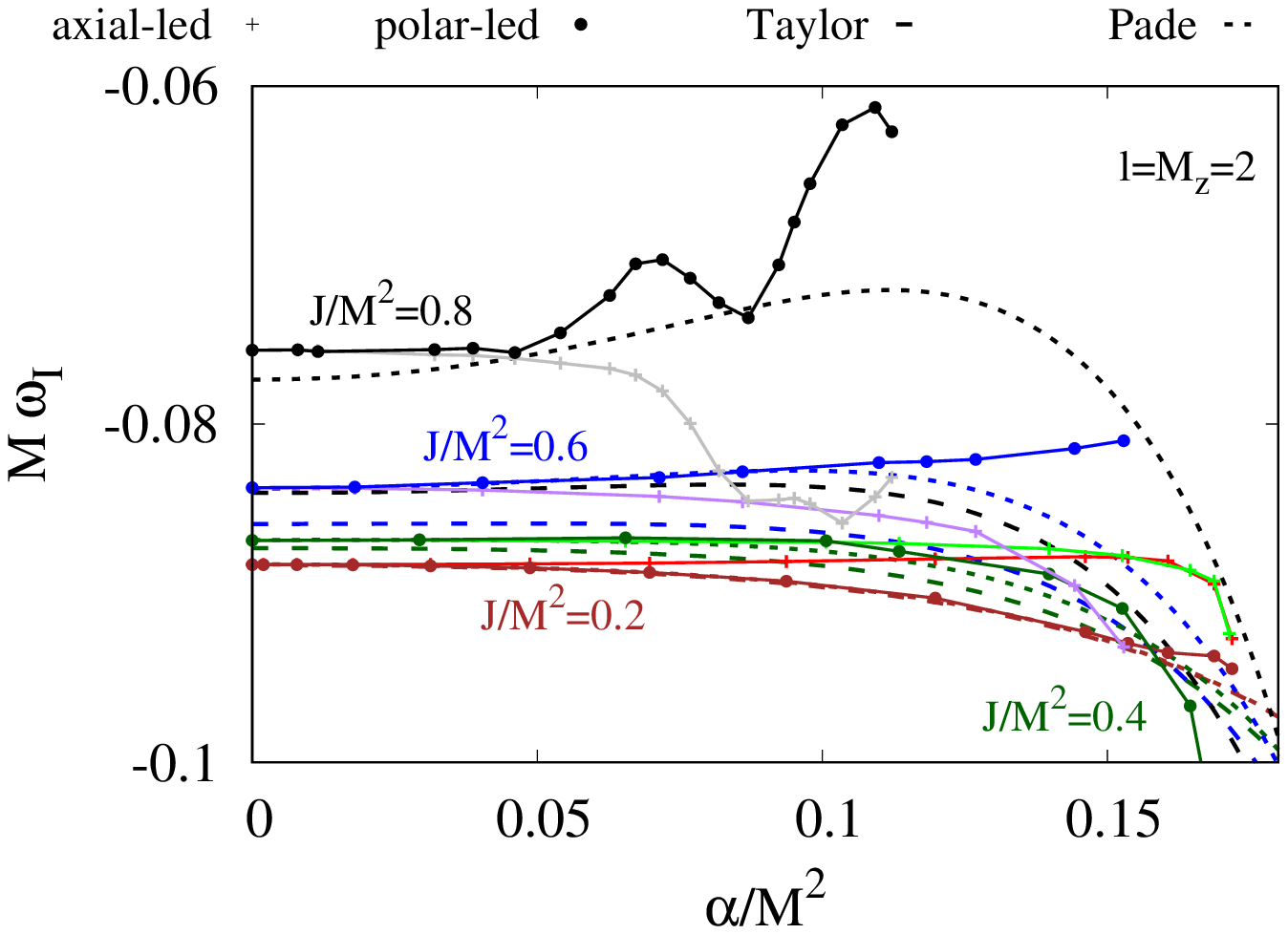}
}
\end{center}
\caption{Comparison of the non-perturbative $M_z=2$, $l=2$-led polar EdGB QNMs
with results obtained to sixth order in the coupling strength $\xi$ and second order
in the rotation parameter $j$ using Taylor and Pad\'e approximations.
\vspace*{1.0cm}}
\label{Fig3}
\end{figure}

We compared the non-perturbative results directly with the slow-rotation and small-coupling approximations.
Figure~\ref{Fig3} shows this comparison for the polar-led $l=2$, $M_z=2$ mode.
For sufficiently small $j$ and $\xi$ the agreement is excellent, as expected.
Toward stronger coupling and more rapid rotation, however, the non-perturbative results become essential.
In particular, the boundary of the domain of existence must be respected.
This illustrates that future precision ringdown tests will require both accurate background solutions and QNMs beyond the perturbative regime.

\subsection{Curvature-induced spontaneous scalarization}

Spontaneous scalarization arises when
\begin{equation}
f'(0)=0,\qquad f''(0)\neq0 .
\end{equation}
Then the GR black holes remain solutions, but they can become tachyonically unstable \cite{Doneva:2017bvd,Silva:2017uqg,Antoniou:2017acq}, when a sufficiently negative effective mass squared arises.
Here, the Gauss-Bonnet invariant acts as an effective mass term.
The simplest such coupling function features only a quadratic coupling
\begin{equation}
f(\varphi)=\frac{\eta}{8}\varphi^2 .
\end{equation}
For Schwarzschild black holes $R^2_{\rm GB}=48M^2/r^6$.
Then only positive coupling can trigger scalarization.
For a pure quadratic coupling the fundamental static scalarized branch is radially unstable, whereas suitable nonlinear coupling functions can yield a radially stable part \cite{Blazquez-Salcedo:2018jnn}.
In contrast, for an exponential coupling \cite{Doneva:2017bvd}, a large portion of the fundamental branch is stable \cite{Blazquez-Salcedo:2018jnn}.

In the presence of rotation, the Gauss-Bonnet term must be examined in the Kerr background,
\begin{equation}
R^2_{\rm GB}
=\frac{48M^2}{(r^2+\chi^2)^6}
\left(r^6-15r^4\chi^2+15r^2\chi^4-\chi^6\right),
\qquad \chi=a\cos\theta .
\end{equation}
This leads to two distinct effects.
For positive coupling, rotation suppresses scalarization \cite{Cunha:2019dwb,Collodel:2019kkx}, while for negative coupling, sufficiently rapid rotation can induce scalarization \cite{Dima:2020yac,Hod:2020jjy,Herdeiro:2020wei,Berti:2020kgk}.
The latter phenomenon is therefore also referred to as \textit{spin-induced scalarization}.

\begin{figure}
\begin{center}
\mbox{ 
\includegraphics[width=0.5\textwidth]{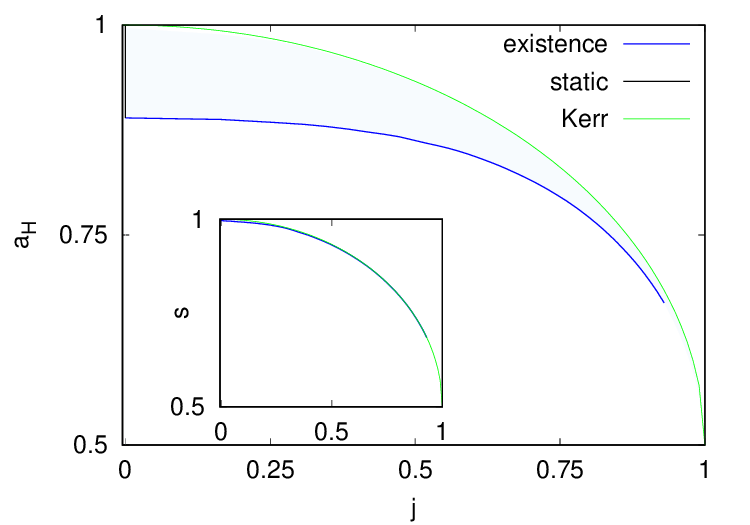}
\includegraphics[width=0.5\textwidth]{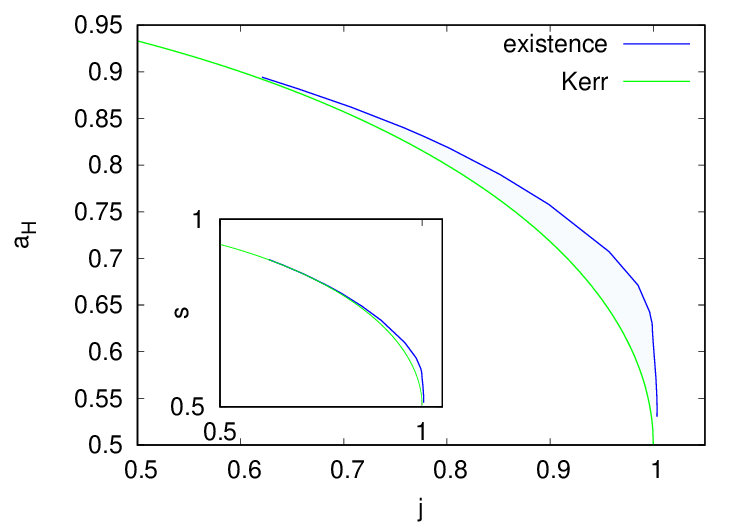}
}
\end{center}
\caption{Domains of existence of rotating EsGB black holes with quadratic
coupling, shown by the dimensionless horizon area $a_{\rm H}=A_{\rm H}/16\pi M^2$
versus $j=J/M^2$: (a) positive coupling, where rotation suppresses
scalarization; (b) negative coupling, where rotation induces scalarization.
The insets show the corresponding dimensionless entropy.}
\label{Fig4}
\end{figure}

We exhibit the domain of existence of rotating EsGB black holes with quadratic coupling in Fig.~\ref{Fig4}.
Here we show the dimensionless horizon area $a_{\rm H}=A_{\rm H}/16\pi M^2$ versus the dimensionless angular momentum $j=J/M^2$ for (a) positive coupling  and (b) negative coupling.
The corresponding dimensionless entropy is seen in the insets.
For positive coupling, the scalarized domain is continuously connected to the static solutions and becomes narrower as the angular momentum increases \cite{Cunha:2019dwb,Collodel:2019kkx}. 
For negative coupling, on the other hand, scalarized solutions emerge only beyond the critical rotation of $j=0.5$ \cite{Dima:2020yac,Hod:2020jjy,Herdeiro:2020wei,Berti:2020kgk}. 
Thus, rotation does not merely deform the scalarized black holes.
It can qualitatively change whether scalarization is possible at all. 
A complete non-perturbative QNM analysis of rapidly rotating spontaneously scalarized black holes remains on our agenda.

\section{Einstein-scalar-Gauss-Bonnet-Ricci black holes}

Spontaneous scalarization assumes that the GR solutions with a vanishing scalar field can be realized cosmologically.
A coupling of the scalar field to the Ricci scalar provides one possibility to make GR a cosmological attractor while retaining curvature-induced black-hole scalarization \cite{Antoniou:2020nax}.
A simple Einstein-scalar-Gauss-Bonnet-Ricci (EsGBR) action is
\begin{equation}
S=\frac{1}{16\pi}\int d^4x\sqrt{-g}
\left[
R-\frac{1}{2}(\partial_\mu\varphi)^2
+\frac{\varphi^2}{2}
\left(\alpha R^2_{\rm GB}-\frac{\beta}{2}R\right)
\right].
\label{esgbr}
\end{equation}
For Schwarzschild black holes, the Ricci scalar vanishes, and therefore the onset of spontaneous scalarization is unchanged by the Ricci coupling. 
The scalarized branches, however, depend strongly on the additional coupling $\beta$ \cite{Antoniou:2021zoy,Antoniou:2022agj}. 
In particular, for sufficiently large $\beta$, a part of the fundamental branch is stable against radial perturbations.

However, radial stability does not necessarily imply a full mode stability. 
Recently, a quadrupole zero mode was found on the fundamental spherical branch that gives rise to new static axially symmetric black holes with prolate and oblate horizons \cite{Kleihaus:2023zzs}. 
Subsequently, a hexadecupole zero mode was uncovered, giving rise to further axially symmetric branches \cite{Kunz:2025ssf}.
In fact, these bifurcations occur for sufficiently large Ricci coupling on the radially stable part of the spherical branch. 
This observation, therefore, provides a clear example where non-radial perturbations reveal an instability that is invisible in the radial sector. 

In our very recent work, we have extended the stability analysis to higher multipoles \cite{Blazquez-Salcedo:2026hierarchy}. 
This study reveals a systematic structure of the angular instabilities beyond the previously known sectors. 
Along a fundamental branch, stability is first lost in the eikonal regime.
This is consistent with the angular Laplacian instability seen in Ref.~\cite{Minamitsuji:2024twp}.
Then, with decreasing mass, a whole hierarchy of angular instabilities emerges with decreasing angular number $l$, extending all the way down to the quadrupole instability.

\section{Conclusions}

Einstein-scalar-Gauss-Bonnet theories provide an attractive arena for strong-gravity tests beyond GR. 
For the dilatonic coupling, GR black holes are no longer solutions and the black holes necessarily carry scalar hair. 
Their rotating solutions possess a restricted domain of existence and can deviate from Kerr in their multipole moments, orbital properties, shadows, and QNM spectra.

The main recent development discussed here is the calculation of the QNM spectrum of rapidly rotating EdGB black holes without expanding perturbatively in the Gauss-Bonnet coupling or in the angular momentum. 
The spectrum contains polar-led, axial-led, and scalar-led families, and the Kerr isospectrality is broken as soon as the coupling is switched on. 
Most importantly for future ringdown observation is that the ordering of the damping times is not fixed but can change considerably.
In particular, the longest-lived mode can change its character completely as the coupling and the spin are varied. 
The comparison with slow-rotation and weak-coupling expansions shows excellent agreement in their regime of validity, while the fully non-perturbative calculation becomes necessary toward stronger coupling and rapid rotation.

For coupling functions allowing spontaneous scalarization, GR black holes remain solutions but can become tachyonically unstable and develop scalar hair. 
Rotation can suppress this mechanism for one sign of the coupling and induce it for the opposite sign. 
For a quadratic coupling, a further Ricci coupling can stabilize a part of the fundamental scalarized branch against radial perturbations. 
However, non-radial instabilities will still occur. 
In fact, a whole hierarchy of angular instabilities arises on the radially stable fundamental branch.
It will be interesting to determine the full QNM spectra of the corresponding rapidly rotating EsGBR black holes and to understand their stability in the presence of rotation. 
Together with increasingly precise gravitational-wave observations, such calculations provide a direct route toward testing the Kerr paradigm and alternative theories of gravity.

\begin{acknowledgments}
We would like to thank all our collaborators for many discussions and for their contributions to the results reviewed here.
\end{acknowledgments}

\section*{Funding}

We gratefully acknowledge support by MICINN project PID2021-125617NB-I00 ``QuasiMode''.
J.L.B.S. gratefully acknowledges support from MICINN project CNS2023-144089 ``Quasinormal modes''.
F.S.K. gratefully acknowledges support from ``Atracci\'on de Talento Investigador Cesar Nombela'' of the Comunidad de Madrid under grant number 2024-T1/COM-31385.

\clearpage


\clearpage

\clearpage

\end{document}